\documentclass[conference]{IEEEtran}

\usepackage{scalerel}
\usepackage{tikz}
\usetikzlibrary{svg.path}

\definecolor{orcidlogocol}{HTML}{A6CE39}
\tikzset{
  orcidlogo/.pic={
    \fill[orcidlogocol] svg{M256,128c0,70.7-57.3,128-128,128C57.3,256,0,198.7,0,128C0,57.3,57.3,0,128,0C198.7,0,256,57.3,256,128z};
    \fill[white] svg{M86.3,186.2H70.9V79.1h15.4v48.4V186.2z}
                 svg{M108.9,79.1h41.6c39.6,0,57,28.3,57,53.6c0,27.5-21.5,53.6-56.8,53.6h-41.8V79.1z M124.3,172.4h24.5c34.9,0,42.9-26.5,42.9-39.7c0-21.5-13.7-39.7-43.7-39.7h-23.7V172.4z}
                 svg{M88.7,56.8c0,5.5-4.5,10.1-10.1,10.1c-5.6,0-10.1-4.6-10.1-10.1c0-5.6,4.5-10.1,10.1-10.1C84.2,46.7,88.7,51.3,88.7,56.8z};
  }
}

\newcommand\orcidicon[1]{\href{https://orcid.org/#1}{\mbox{\scalerel*{
\begin{tikzpicture}[yscale=-1,transform shape]
\pic{orcidlogo};
\end{tikzpicture}
}{|}}}}

\usepackage{cite}
\usepackage{amsmath,amssymb,amsfonts}
\usepackage{algorithmic}
\usepackage{graphicx}
\usepackage{textcomp}
\usepackage{xcolor}
\usepackage{siunitx}
\usepackage{hyperref}
\usepackage[export]{adjustbox}
\usepackage{booktabs}
\usepackage{multirow}

\makeatletter

\usepackage{eso-pic}

\def\BibTeX{{\rm B\kern-.05em{\sc i\kern-.025em b}\kern-.08em
    T\kern-.1667em\lower.7ex\hbox{E}\kern-.125emX}}
\begin{document}

\title{Deep Learning-based Symbolic Indoor Positioning using the Serving eNodeB\\}

\author{\IEEEauthorblockN{Fahad Alhomayani \orcidicon{0000-0002-4914-8722} and Mohammad Mahoor}
\IEEEauthorblockA{\small Department of Electrical and Computer Engineering, University of Denver, USA \\
\texttt{\href{mailto:fahad.al-homayani@du.edu}{fahad.al-homayani@du.edu}} and \texttt{\href{mailto:mmahoor@du.edu}{mmahoor@du.edu}}}}

\maketitle

\begin{abstract}
This paper presents a novel indoor positioning method designed for residential apartments. The proposed method makes use of cellular signals emitting from a serving eNodeB which eliminates the need for specialized positioning infrastructure. Additionally, it utilizes Denoising Autoencoders to mitigate the effects of cellular signal loss. We evaluated the proposed method using real-world data collected from two different smartphones inside a representative apartment of eight symbolic spaces. Experimental results verify that the proposed method outperforms conventional symbolic indoor positioning techniques in various performance metrics. To promote reproducibility and foster new research efforts, we made all the data and codes associated with this work publicly available.
\end{abstract}

\begin{IEEEkeywords}
Cellular Networks, Dataset, Deep Learning, Denoising Autoencoder, eNodeB, Symbolic Indoor Positioning.
\end{IEEEkeywords}

\section{Introduction}

Interest in indoor positioning research has grown substantially in recent years. This can be attributed to the multitude of potential applications enabled by indoor positioning \cite{6803131,6399501,werner2014indoor,ferreira2017localization}. Yet, designing an indoor positioning system has remained a challenging task since indoor environments are very complex and are often characterized by non-line-of-sight settings, moving people and furniture, walls of different densities, and the presence of different indoor appliances that alter indoor signal propagation. Among the techniques used for indoor positioning, location fingerprinting, or simply fingerprinting, has received the most attention due to its simplicity and ability to produce accurate positioning estimates \cite{doi:10.1080/17489725.2020.1817582}. The concept of fingerprinting is to identify indoor spatial locations based on location-dependent measurable features (i.e., location fingerprints). Examples of location fingerprints include radio frequency fingerprints (WiFi \cite{nowicki2017low}, Bluetooth \cite{7103024}, cellular \cite{8570849}), magnetic field fingerprints \cite{8626558}, image fingerprints \cite{7743683}, and hybrid fingerprints \cite{doi:10.1080/17489725.2018.1455992}. However, the main drawback of fingerprinting is the laborious and time-consuming site surveying task in which fingerprints are collected at predefined reference points (RPs) with known coordinates. Depending on the area to be covered by the system and the accuracy requirement, the number of required RPs can be significant. Symbolic positioning tries to relax this requirement by collecting fingerprints in zones rather than at points \cite{doi:10.1080/17489725.2018.1455992}. However, the concept of distance is lost since zones are independent and the user’s location is now expressed symbolically (e.g., “in the kitchen”) instead of physically (using a coordinate system).

\begin{figure}[!t]
\centering
\includegraphics[width=0.5\columnwidth]{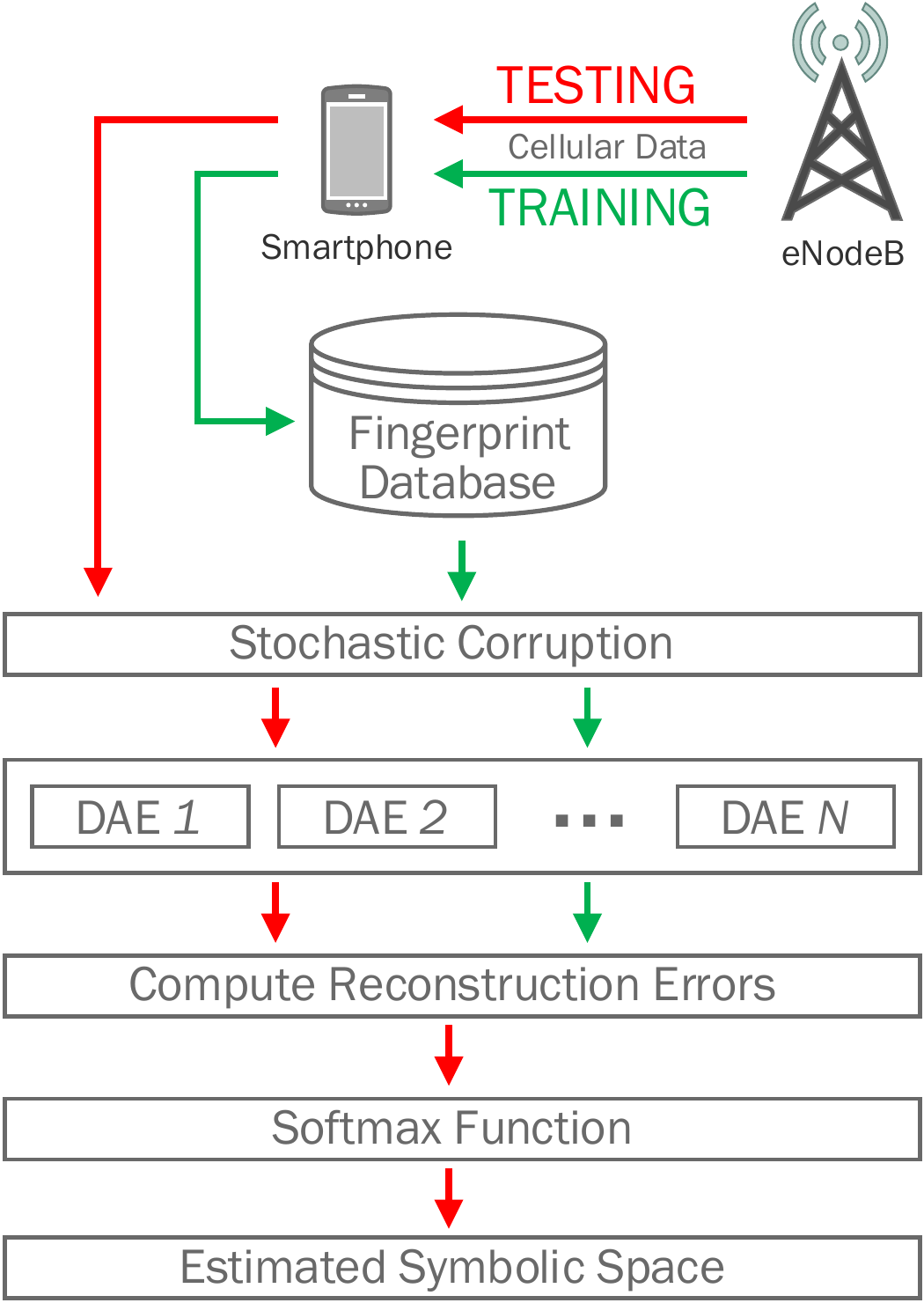}
\caption{The general scheme of the proposed method representing the training and testing phases.}
\label{Scheme}
\end{figure}

In this paper, we treat the indoor positioning problem as a classification problem. Each symbolic space in the environment has different cellular signal propagation characteristics and, hence, should be considered as a unique class. To distinguish one class from another (i.e., one symbolic space from another), we leverage Denoising Autoencoders (DAEs). The motivation behind employing DAEs, as opposed to other learning algorithms, is their ability to handle noisy data effectively and efficiently. Our experimental results, which are based on real signal measurements collected inside a residential apartment, verify that the proposed method outperforms conventional symbolic indoor positioning techniques on various performance metrics. The general scheme of the proposed method is depicted in Fig. \ref{Scheme}.

The remainder of this paper is organized as follows. Section \hyperref[sec2]{II} reviews some of the recent work in deep learning-based indoor positioning. Section \hyperref[sec3]{III} describes and validates the dataset used in this study. Section \hyperref[sec4]{IV} provides background on Autoencoders and discusses the design of the proposed method. Section \hyperref[sec5]{V} reports on the evaluation experiments and analyzes the results. Section \hyperref[sec6]{VI} concludes the paper.

\section{Related Work}
\label{sec2}
In this section, a review of some recent research efforts that utilize machine learning for symbolic indoor positioning is provided, followed by a review of some recent research efforts that utilize machine learning for cellular-based indoor positioning.

\subsection{Machine Learning for Symbolic Indoor Positioning}

Werner \textit{et al}.\cite{7743683} utilized the Convolutional Neural Network (CNN)-based AlexNet as a generic feature extractor to classify a query image to one of \num{16} rooms. No fine-tuning was performed on the pre-trained network; instead, the authors directly fed the features extracted by the first Fully-Connected (FC) layer to a Naïve Bayes classifier. These features helped their model to generalize from local to global views (i.e., from small views in training to large views in testing) well. However, this did not hold when attempting to generalize from global to local views. Nowicki and Wietrzykowski \cite{nowicki2017low} used an Autoencoder (AE) followed by an FC network for multi-building and multi-floor classification using WiFi fingerprints. The purpose of the AE is to perform dimensionality reduction. This is important because a WiFi fingerprint has entries for all Access Points (APs) detected in an entire environment, but only a subset of these APs is observed for different locations. This is especially true for large-scale environments. Most recently, Tamas and Toth \cite{doi:10.1080/17489725.2018.1455992} performed a performance analysis of five machine learning classifiers for symbolic indoor positioning. They used hybrid fingerprints (WiFi, Bluetooth, and magnetometer) to evaluate and compare the classifiers being studied. The fingerprints were obtained from the Miskolc IIS dataset which contains measurements from \num{21} zones of different sizes inside a three-story university building.

Our proposed method has several advantages compared to the aforementioned works:

\begin{itemize}
\item It preserves privacy because it does not require the capturing of images for positioning.
\item It is well-suited for small-scale indoor environments, namely residential apartments and homes, where people spend most of their time.
\item It does not require on-premises infrastructures such as WiFi APs or Bluetooth beacons for operation. Instead, it relies on omnipresent cellular signals.
\item It has little overhead because only a single fingerprint type is required for positioning which eliminates the complexity associated with fusing multiple fingerprint types. 
\end{itemize}

\subsection{Machine Learning for Cellular-based Indoor Positioning}

Rizk \textit{et al}. \cite{8570849} used an FC network to perform cellular Received Signal Strength (RSS) fingerprinting. Data augmentation techniques were used to increase the training set by \num{8}-fold. The authors achieved a positioning error of less than \SI{3}{\meter} \SI{90}{\percent} of the time. However, to achieve this accuracy, the RSS from \num{17} Second-Generation (\num{2}G) cellular Base Stations (BSs) had to be measured. Later, the authors used a Recurrent Neural Network (RNN) to capture the temporal dependency between consecutive RSS measurements received from the serving BS \cite{rizk2019monodcell}. The achieved positioning accuracy was comparable to that acquired by their previous approach, however, a measurement window of at least three seconds had to be fed to the RNN. Arnold \textit{et al}. \cite{8446013} used a custom-built linear array of Multiple-Input Multiple-Output (MIMO) antennas installed in a \SI{20}{\meter} by \SI{7}{\meter} area for indoor positioning. They used an FC network to correlate the antennas' channel coefficients to a \num{3}D position relative to the array's location. Vieira \textit{et al}. \cite{8292280} used a CNN to learn the structure of massive MIMO channels for indoor positioning. A cellular channel model was used to generate unique channel fingerprints for each training/testing position. These fingerprints represent clusters of multipath components obtained from a BS equipped with a linear array of antennas. Since all measurements were based on simulated data, real-world measuring impairments such as noise and channel fading were not considered.

Compared to the previous works, our proposed method combines several features that place it in a unique position:
\begin{itemize}
\item It employs DAEs to handle incomplete measurements caused by unpredictable cellular signal loss.
\item It only utilizes the information measured from the serving BS, which is a Fourth-Generation (\num{4}G) cellular BS (also known as an eNodeB in Long-Term Evolution).
\item It is well-suited for real-time positioning applications, given the parametric nature of DAEs, in addition to performing one-shot positioning (i.e., only a single fingerprint sample is required to estimate the user’s location). 
\item It is based on real-world measurements emitting from a real eNodeB. No simulated, interpolated, or augmented data were used in this study. 
\item It is well-suited for smartphone-based positioning because all measurements were collected using smartphones as opposed to custom-built collection platforms.
\end{itemize}

\section{Dataset Description}
\label{sec3}
Nearly all indoor positioning solutions found in the literature were evaluated using private data. Thus, the results obtained are self-reported and cannot be reproduced. Additionally, the lack of publicly available datasets that can be used to develop, evaluate, and compare indoor positioning solutions constitutes a high entry barrier for studies. For these reasons, we made the dataset used in this study publicly available \cite{alhomayani_mahoor_2020}. The following subsections describe the data collection platform, environment, procedure, and technical quality, respectively.

\subsection{Data Collection Platform}
\begin{figure}[!t]
    \centering
    \includegraphics[width=0.3\columnwidth, fbox]{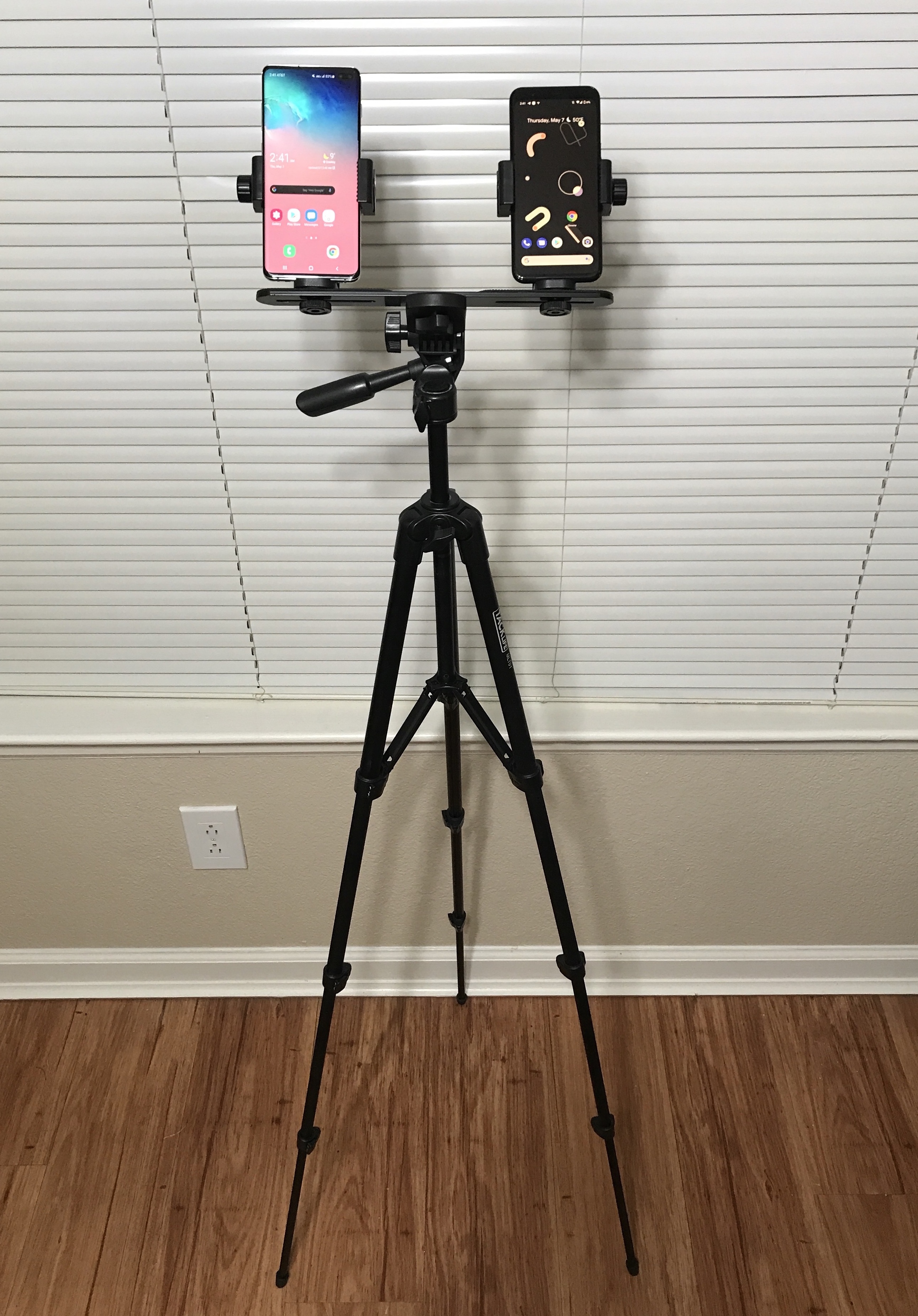}
    \caption{A picture of Phone 1 and Phone 2 attached to the tripod}
    \label{phones}
\end{figure}

We used two smartphones for data acquisition: Samsung’s Galaxy S\num{10}+ (Phone \num{1}) and Google’s Pixel \num{4} (Phone \num{2}). Both smartphones ran on Android \num{10}. The motivation behind choosing Android-powered smartphones was twofold. First, Android provides Application Programming Interfaces (APIs) that allow for acquiring raw data at the hardware level. Second, Android-powered smartphones account for over \SI{74}{\percent} of the market share worldwide \cite{statcounter}. We attached the two smartphones to a tripod using a dual mount (Fig. \ref{phones}). Both smartphones were in portrait mode and were kept at a fixed height of \SI{130}{\cm}. The tripod head was adjusted to tilt the smartphones at a $\sim$\num{40}-degree angle to the vertical plane. We installed the same third-party app \cite{NetMonitorPro} used for the data collection on both smartphones. The app allowed for conveniently collecting and exporting cellular network data.

\begin{figure}[!b]
\centering
\includegraphics[width=0.9\columnwidth]{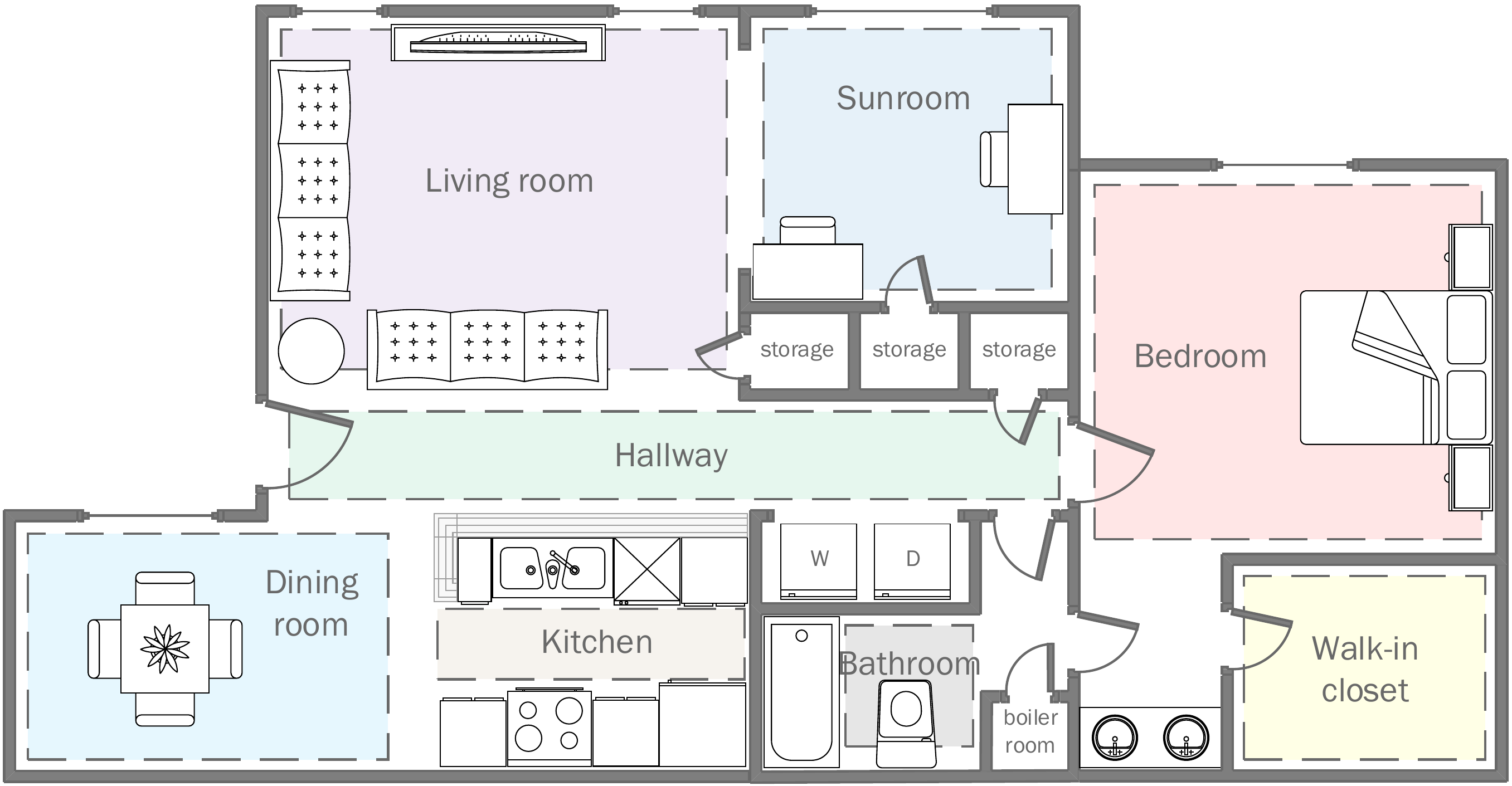}
\caption{Layout of the apartment where data was collected}
\label{Layout}
\end{figure}

\subsection{Data Collection Environment}
We performed data collection inside a residential apartment of eight symbolic spaces. As seen from the apartment’s layout (Fig. \ref{Layout}), the symbolic spaces include a living room (\num{4.0}$\times$\SI{3.0}{\meter\squared}), a sunroom (\num{2.6}$\times$\SI{2.3}{\meter\squared}), a bedroom (\num{3.5}$\times$\SI{3.2}{\meter\squared}), a hallway (\num{7.0}$\times$\SI{0.8}{\meter\squared}), a dining room (\num{3.2}$\times$\SI{2.0}{\meter\squared}), a kitchen (\num{2.8}$\times$\SI{0.6}{\meter\squared}), a bathroom (\num{1.1}$\times$\SI{1.1}{\meter\squared}), and a walk-in closet (\num{2.2}$\times$\SI{1.6}{\meter\squared}). The floor plan delineating the apartment’s dimensions is provided alongside the dataset in the form of a \texttt{.vsdx} file.

\subsection{Data Collection Procedure}
\begin{figure}[!t]
\centering
\includegraphics[width=0.9\columnwidth]{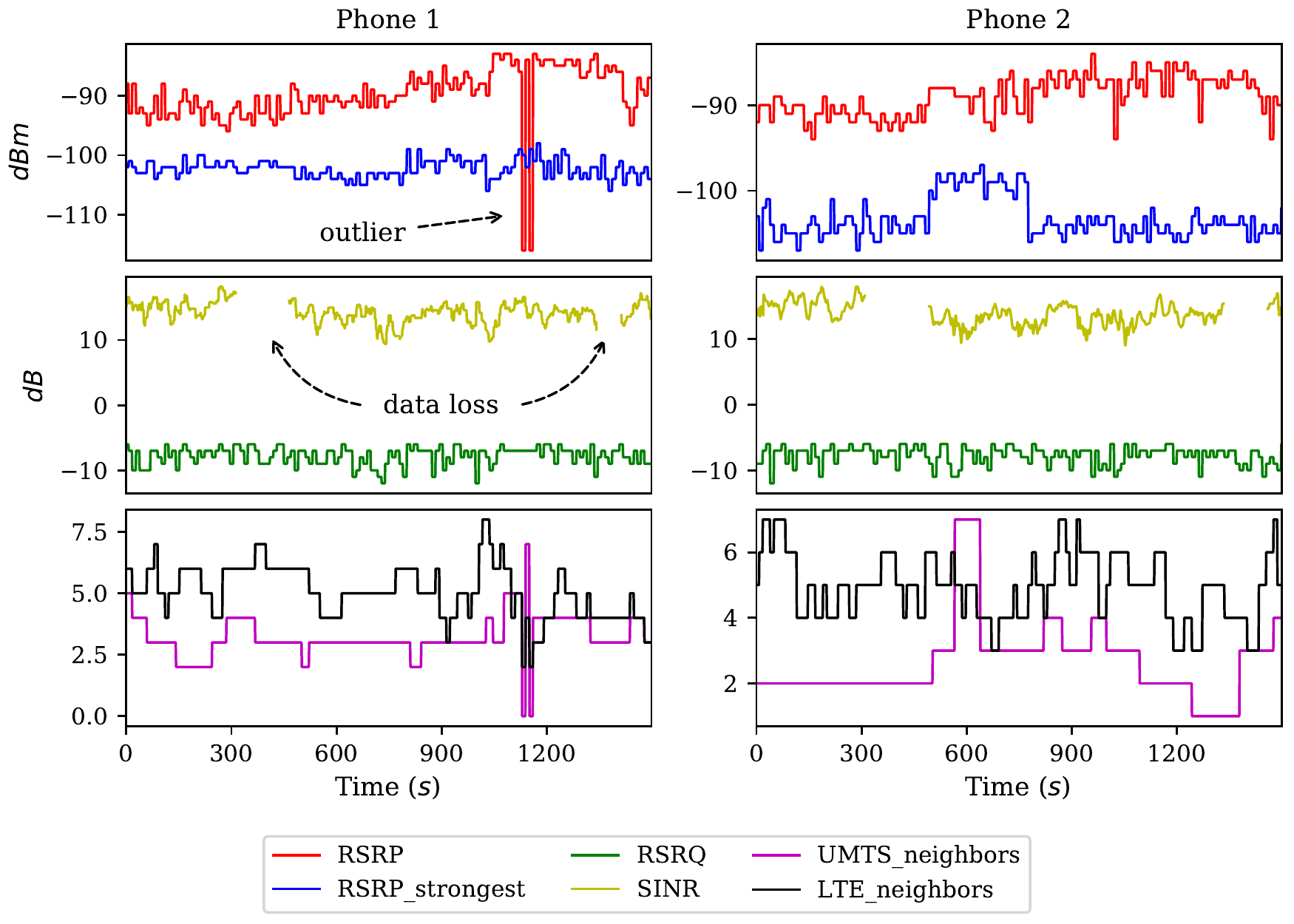}
\caption{Plots of cellular data showing examples of outliers and data loss in the data collected by Phone \num{1} and Phone \num{2} inside the walk-in closet.}
\label{Walk-in_closet}
\end{figure}

A smartphone’s cellular modem constantly scans the cellular network for cell selection/reselection and handover purposes. Android provides APIs to extract data associated with scans such as Reference Signal Received Power (RSRP) and cell identity information \cite{android_telephony}. For each of the symbolic spaces described above, we collected \num{25} minutes of cellular data (per phone) at a sampling rate of \num{1} hertz (\si{\hertz}). During data collection, we systematically changed the position and orientation of the tripod to uniformly cover space and direction. Sampling results were exported as a \texttt{.csv} file and named with the smartphone’s and space’s name  (e.g., \texttt{Phone2\_Bedroom.csv}). Table \ref{table1} lists all fields in each data sample and their descriptions. As an example, Fig. \ref{Walk-in_closet} plots the data collected from the smartphones located inside the walk-in closet.

\begin{table*}[!h]
\caption{Field labels of data samples and their description}
\scriptsize
\centering
\begin{adjustbox}{width=0.8\width,center=\width}
\resizebox{\width}{!}{
\begin{tabular}{@{}m{0.2cm}m{2.5cm}m{13.5cm}@{}} 
\toprule
\bfseries \# & \bfseries Field label & \bfseries Description \\  
\midrule
1 & \texttt{Date\_Time} & The date and time the sample was captured as \texttt{YYYYMMDDhhmmss} \\ 
2 & \texttt{PLMN\_ID} & The Public Land Mobile
Network (PLMN) IDentifier\\
3 &   \texttt{eNodeB\_ID} & The E-UTRAN NodeB (eNodeB) IDentifier that is used to uniquely identify an eNodeB within a PLMN \\
4 &  \texttt{Cell\_ID} & The Cell IDentifier which is an internal descriptor for a cell. It can take any value between \texttt{0} and \texttt{255} \\ 
5 &  \texttt{ECI} & The E-UTRAN Cell Identifier that is used to uniquely identify a cell within a PLMN. \texttt{ECI} = \num{256} $\times$ \texttt{eNodeB\_ID} + \texttt{Cell\_ID}. \\
6 &   \texttt{RSRP} & The Reference Signal Received Power in decibel-milliwatts (\si{dBm})  \\ 
7 &    \texttt{RSRQ} & The Reference Signal Received Quality in decibels (\si{dB})\\
8 & \texttt{SINR} & The Signal to Interference and Noise Ratio in \si{dB}\\ 
9 & \texttt{UMTS\_neighbors} & The number of neighboring Universal Mobile Telecommunications Service (UMTS) cells \\ 
10 & \texttt{LTE\_neighbors} & The number of neighboring Long-Term Evolution (LTE) cells \\ 
11 &  \texttt{RSRP\_strongest} & The Reference Signal Received Power, in \si{dBm}, corresponding to the strongest neighboring LTE cell \\ 

\bottomrule
\end{tabular}
}
\end{adjustbox}
\label{table1}
\end{table*}

\subsection{Technical Validation}
The technical quality of the dataset was evaluated using experiments
that consider its reliability and validity:

\subsubsection{Measurement Reliability}
Before the collection campaign, we captured cellular data over three different days at the same location. We used Spearman’s
and Kendall’s correlation coefficients to quantify the degree of consistency between temporal measurements for a given phone. Table \ref{table2} shows Spearman's and Kendall's correlation coefficients for the two smartphones for all possible pairs of days. Given that correlation results are high (i.e., close to the maximum value of \num{1.0}), it can be concluded that the dataset possesses a high degree of reliability.

\subsubsection{Measurement Validity} We assessed measurement validity by comparing the cellular data captured by the two phones and checking for consistency. Accordingly, for a given day, we used Spearman's and Kendall's correlation coefficients to quantify the degree of consistency between the measurements obtained by the phones. The correlation results for the foregoing three days are shown in Table \ref{table3}. These results demonstrate high levels of consistency, which attests to the validity of the dataset.

\begin{table*}[!h]
\caption{Results of the correlation analysis between the measurements obtained on three different days for Phone \num{1} and Phone \num{2}. The results were generated using synchronized readings of fields \num{6}--\num{11}. The \textit{p}-values of all results were less than \num{0.01}.} 
\begin{adjustbox}{width=0.62\textwidth,center=\textwidth}
\resizebox{\textwidth}{!}{
\centering
	\begin{tabular}{@{}r*{6}{S[table-format=-3.4]}@{}}
	\toprule
	& \multicolumn{3}{c}{\bfseries Phone 1} & \multicolumn{3}{c}{\bfseries Phone 2} \\
	\cmidrule(lr){2-4} \cmidrule(lr){5-7}
	& {$\{Day1,Day2\}$} & {$\{Day2,Day3\}$} & {$\{Day1,Day3\}$} & {$\{Day1,Day2\}$} & {$\{Day2,Day3\}$} & {$\{Day1,Day3\}$} \\
	\cmidrule(lr){2-2} \cmidrule(lr){3-3} \cmidrule(lr){4-4} \cmidrule(lr){5-5} \cmidrule(lr){6-6} \cmidrule(lr){7-7} 
	\textit{Spearman's} $\rho$ & 0.992 & 0.988 & 0.981 &  0.990 & 0.976 & 0.980  \\  
	\textit{Kendall's} $\tau$ & 0.983 & 0.973 & 0.956 &  0.977 & 0.945 & 0.953  \\ 
	\bottomrule  
\end{tabular} 
} 
\end{adjustbox}
\label{table2}
\end{table*}

\begin{table}[!h]
\caption{Results of the correlation analysis between the measurements obtained from Phone \num{1} \& Phone \num{2} for three different days. The results were generated using readings of fields \num{6}--\num{11}. The \textit{p}-values of all results were less than \num{0.02}.} 
\begin{adjustbox}{width=0.6\columnwidth,center=\columnwidth}
\resizebox{\columnwidth}{!}{
\centering
	\begin{tabular}{@{}r*{3}{S[table-format=-3.4]}@{}}
	\toprule
	& \multicolumn{3}{c}{\bfseries Phone 1 \& Phone 2} \\
	\cmidrule(lr){2-4}
	& {$Day1$} & {$Day2$} &  {$Day3$}\\
	\cmidrule(lr){2-2} \cmidrule(lr){3-3} \cmidrule(lr){4-4}
	\textit{Spearman's} $\rho$ & 0.991 & 0.995 & 0.968 \\  
	\textit{Kendall's} $\tau$ & 0.979 & 0.989 & 0.927 \\      
	\bottomrule  
\end{tabular} 
}
\end{adjustbox}

\label{table3}
\end{table}

\section{Background and Proposed Method}
\label{sec4}

\subsection{Autoencoders}
\begin{figure}[!t]
\centering
\includegraphics[width=0.7\columnwidth]{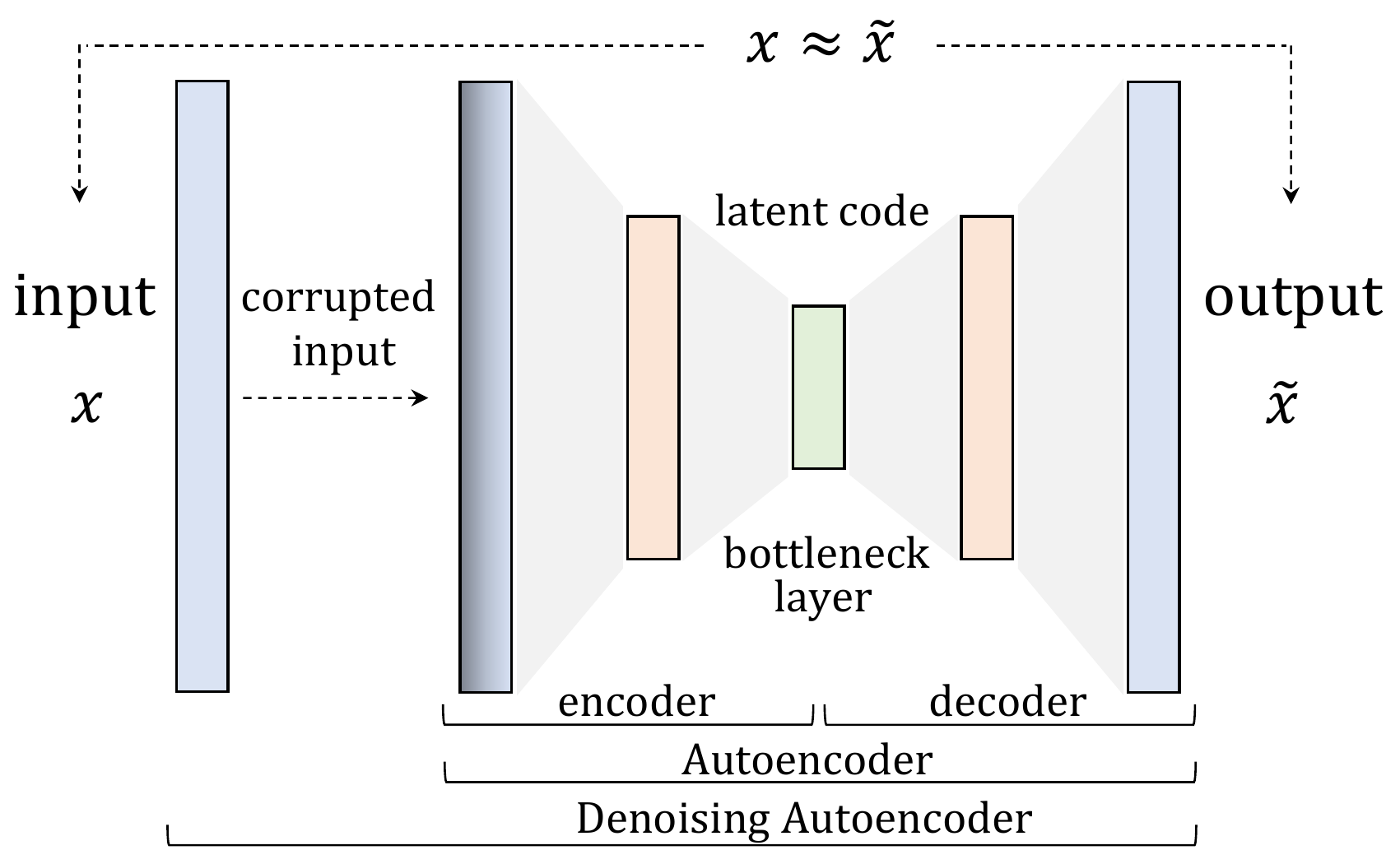}
\caption{The structure of an AE and a DAE}
\label{DAE}
\end{figure}

Autoencoders (AEs) are a family of feedforward neural networks that have been used in unsupervised learning tasks. AEs have the same number of neurons in the input layer as the output layer. A typical AE is trained to reconstruct an input without memorizing or directly copying it. Instead, an encoder-decoder approach is used, as seen in Fig. \ref{DAE}. This hourglass-shaped architecture forces the network to encode (compress) the input into a latent code from which the input can be decoded (reconstructed). Backpropagation is used to learn the network's parameters by minimizing reconstruction loss between the input and the reconstructed input. Common variants of AEs are Denoising AEs (DAEs). DAEs are trained to reconstruct an input from a corrupted version of it (Fig. \ref{DAE}).

\subsection{Proposed Method}
The design of the proposed method is inspired by the successful application of AEs for anomaly detection \cite{chalapathy2019deep}: An AE, when solely trained on normal data instances, fails to reconstruct abnormal data instances, hence, producing a large reconstruction error. The data instances that produce high residual errors are considered outliers. 

The proposed method takes a normalized cellular data sample captured from the serving eNodeB as input \eqref{eq1} and produces an output \eqref{eq2} that is a probability distribution over the set of symbolic spaces in the environment: 

\begin{footnotesize}
\begin{equation}
\begin{split}
X=\big[\texttt{RSRP},\texttt{RSRQ},\texttt{SINR},\texttt{UMTS\_neighbors},\\  \texttt{LTE\_neighbors},\texttt{RSRP\_strongest}\big]; \\  X \in  \mathbb{R}^6: \{x_i \in \mathbb{R} \mid 0 \leq x_i \leq 1\}
\end{split}
\label{eq1}
\end{equation}
\end{footnotesize}

\begin{footnotesize}
\begin{equation}
\begin{split}
Y=\big[\Pr(space_1 \mid X),\cdots,\Pr(space_n \mid X)\big]; \\ Y \in \mathbb{R}^n: \{y_i \in \mathbb{R}  \mid y_1+\cdots+y_n=1\}
\end{split}
\label{eq2}
\end{equation}
\end{footnotesize}

Given DAEs' data-driven learning ability, the proposed method does not make any assumptions about feature independence or the nature of the boundary separating the classes. 

The input vector is corrupted to emulate a randomized loss of cellular data. This is accomplished using a Hadamard product of \eqref{eq1} and an all-ones vector ($\vec{1}$) whose elements are randomly set to \num{0} with a predefined probability $p_{loss}$. For example, if $p_{loss}$ is set to \num{0.5}, there is a \SI{50}{\percent} chance that a given field entry will be set to zero.

During the training phase, a dedicated DAE is employed for each symbolic space. Each DAE is solely trained on the data collected at its corresponding symbolic space. By following this training strategy, we expect that, during the testing phase, all DAEs, except for one, will generate a relatively high reconstruction error when fed the same testing sample. Consequently, the symbolic space associated with the DAE that generated the lowest reconstruction error is considered as the estimated symbolic space. To construct \eqref{eq2}, we used a Softmax function \eqref{eq3} during the testing phase:

\begin{footnotesize}
\begin{equation} 
\label{eq3} 
\Pr(space_i \mid X)=S(\mathcal{L}_i) = \frac{\exp(1/\mathcal{L}_i)}{\sum_{i=1}^{n}\exp(1/\mathcal{L}_i)}
\end{equation}
\end{footnotesize}
where $\mathcal{L}_i$ is the reconstruction loss generated by the $i^{th}$ DAE.

\subsubsection{DAE Architecture}
All DAEs have the same architecture which consists of an input layer of \num{256} neurons, a hidden layer (and its mirror layer) of \num{64} neurons, a bottleneck layer of \num{16} neurons, and an output layer of \num{256} neurons. We developed the DAEs using Keras with the hyperparameters listed in Table \ref{table4}. We selected these hyperparameters using grid search and cross-validation. We used early stopping and dropout to avoid overfitting. 

\begin{table}[!t]
\caption{Hyperparameter tuning}
\scriptsize
\begin{adjustbox}{width=0.6\columnwidth,center=\columnwidth}
\resizebox{\columnwidth}{!}{
\centering
\begin{tabular}{@{}m{2cm}m{3.6cm}@{}} 
\toprule
\bfseries Hyperparameter & \bfseries Value \\  
\midrule
Batch size & \num{100}  \\ 
Dropout rate& \num{0.1} \\ 
Optimizer  & Adadelta ($\rho=\num{0.95}$, $\epsilon=\num{1e-7}$)  \\ 
Learning rate & \num{1.0} \\ 
Activation function  & ReLU \\
Epochs & \si{1,200} \\
Loss function & Binary cross-entropy \\
Weight initializer & Xavier uniform  \\
Bias initializer & Zeros \\
\bottomrule
\end{tabular}
}
\end{adjustbox}
\label{table4}
\end{table}

\subsubsection{Dataset Preprocessing}
From each symbolic space, there were \si{1,500} samples collected by each smartphone. For the entire collection period, and throughout the collection environment, Phone \num{1} and Phone \num{2} camped on the same LTE cell (i.e., \texttt{ECI:98059528}). Thus, entries for field labels \num{2}--\num{5} were identical for all samples. For a given symbolic space, we combined the samples collected by Phone \num{1} and Phone \num{2} to create a single dataset for training and testing the corresponding DAE. After the samples in the combined dataset were randomly shuffled, we allocated \SI{80}{\percent} of them for training and validation, and the remaining \SI{20}{\percent} for testing. Since input features are measured in different units, their values were normalized between \num{0} and \num{1}. This was performed after the dataset was split to avoid data contamination. Fig. \ref{Scheme} shows the general scheme of the proposed method.

\section{Experiments and Results}
\label{sec5}

This section evaluates the performance of the proposed method and investigates the impact of $p_{loss}$ and device heterogeneity on positioning accuracy. Associated computing scripts are publicly available in our figshare repository \cite{alhomayani_mahoor_2020}.

\begin{figure*}[!t]
\centering
\includegraphics[width=0.9\textwidth]{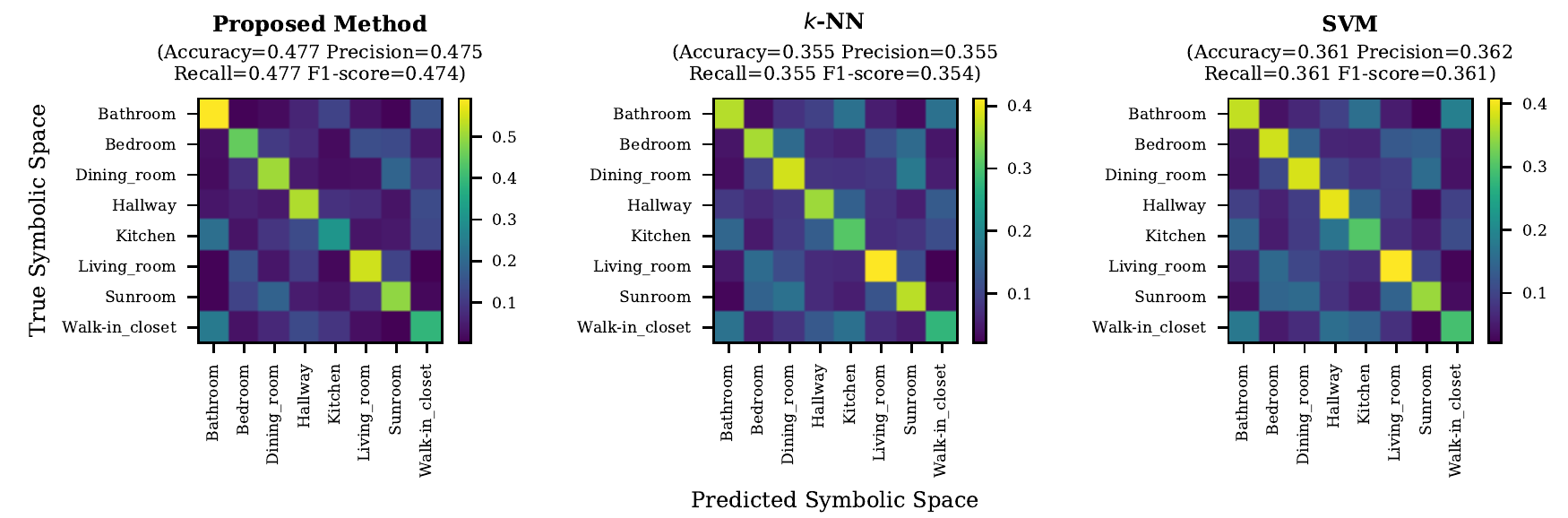}
\caption{Accuracy, Precision, Recall, F\num{1}-score, and the confusion matrix of symbolic space prediction of the proposed method, $k$-NN, and SVM.}
\label{cm}
\end{figure*}

\subsection{Performance Evaluation}
We trained the proposed method with a $p_{loss}$ value of \num{0.5} applied to the training set. The metrics used for performance evaluation are Accuracy, Precision, Recall, and F\num{1}-score as defined in \cite{sklearn_metrics}. We compared the performance of the proposed method against classifiers that are extensively used for indoor positioning, namely $k$-Nearest Neighbor ($k$-NN) and Support Vector Machine (SVM).  For the sake of fair comparison, we trained these classifiers on the same training set used for the proposed method and fine-tuned their parameters using grid search and cross-validation. The testing set used for comparison was contaminated with a $p_{loss}$ value of \num{0.5}. Fig. \ref{cm} reports on the classification results and shows the confusion matrices of the three methods. The results clearly show that the proposed method outperforms both $k$-NN and SVM on all metrics. As mentioned earlier, both smartphones connected to the same LTE cell throughout the environment. However, it is possible, depending on network parameters, that a connection alternates between multiple cells. Incorporating the information obtained by additional cells is expected to further enhance performance because location discernibility will increase with increased features.

Interesting observations can be made by examining the confusion matrices in Fig. \ref{cm}. For example, higher degrees of confusion tend to occur between symbolic spaces that are close to each other (e.g., Kitchen and Bathroom or Sunroom and Living room). Nevertheless, observations exist that prove contrary to this assumption. For instance, there is low confusion between Bedroom and Walk-in closet despite their proximity. In fact, it is more likely to confuse Bedroom for Dining room than it is to confuse Bedroom for Walk-in closet. Such observations could be the result of the complex changes that cellular signals undergo when propagating indoors. Confirming this conjecture is a topic of future research.

\subsection{Effect of $p_{loss}$ on Accuracy}

\begin{figure}[!t]
\centering
\includegraphics[width=0.9\columnwidth]{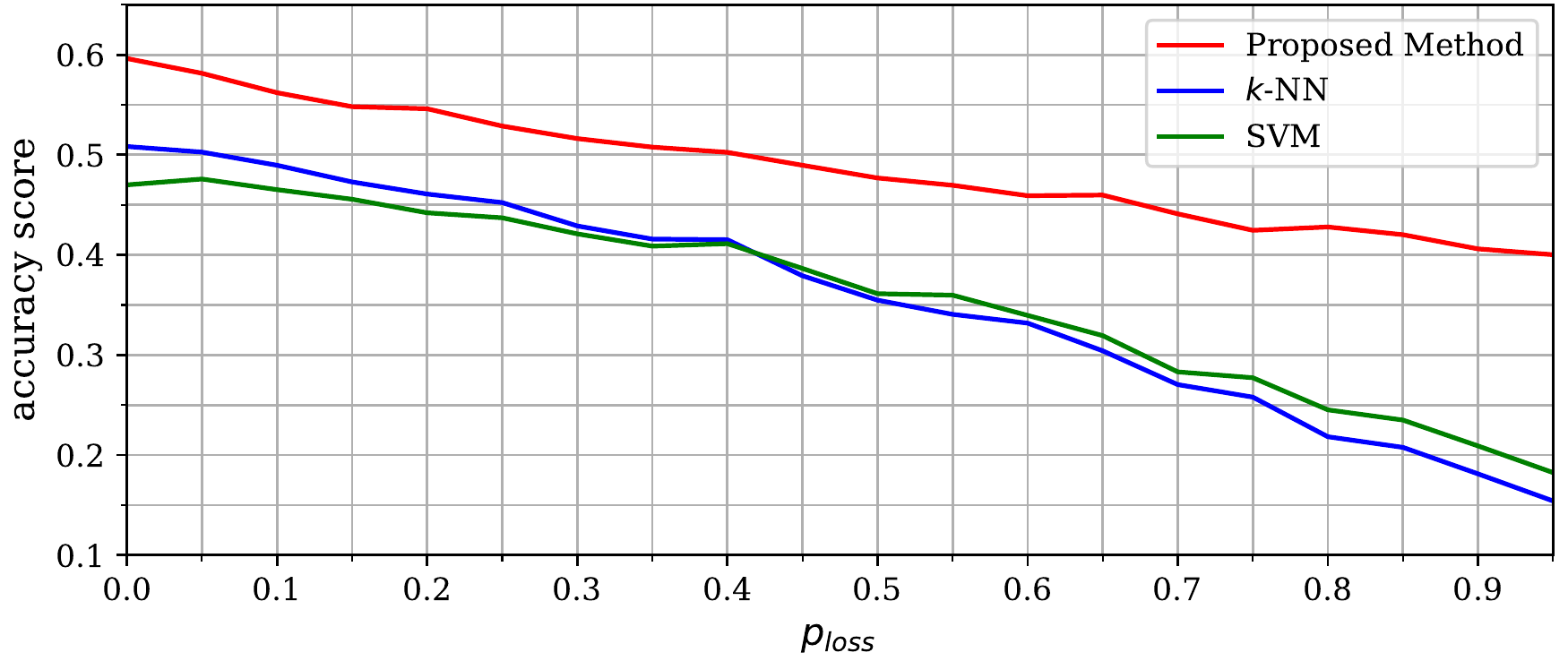}
\caption{The effect of $p_{loss}$ on accuracy for the proposed method, $k$-NN, and SVM.}
\label{avc}
\end{figure}

To study the impact of $p_{loss}$ on positioning accuracy, we evaluated the proposed method, $k$-NN, and SVM using data contaminated with varying $p_{loss}$ values. More specifically, we generated \num{20} copies of the testing set and contaminated each copy with a different $p_{loss}$ value that ranged from \num{0.0} to \num{0.95}, using \num{0.05} increments. The methods' accuracy scores that corresponded to each $p_{loss}$ value were recorded and plotted in Fig. \ref{avc}. One observation that can be made from Fig. \ref{avc} is that, as loss increases, so does the performance gap between the proposed method and $k$-NN/SVM. This primarily suggests that DAEs learned more robust features than $k$-NN and SVM.

\subsection{Effect of Device Heterogeneity on Accuracy}
Smartphones obtain cellular data readings from their cellular chipsets. Since these chipsets are manufactured by different vendors, hardware and firmware specifications are not uniform across smartphones. This results in heterogeneous reception characteristics which, in turn, can degrade the accuracy of the positioning system \cite{7874080}. 

In this experiment, we investigated device heterogeneity by training the proposed method, $k$-NN, and SVM on the data obtained from one smartphone and testing on the data obtained from 1) the same smartphone and 2) the other smartphone to quantify the difference in performance. Table \ref{tableDH} reports on the experiment’s results. As clearly seen from Table \ref{tableDH}, device heterogeneity is a significant problem in all three methods. Average accuracy drops of \SI{45.7}{\percent},  \SI{40.8}{\percent}, and \SI{42.9}{\percent} are observed in the proposed method, $k$-NN, and SVM, respectively. In the field of indoor positioning, there is ongoing research regarding overcoming device heterogeneity and we intend to address this limitation in subsequent research.

\begin{table}[!h]
\caption{Results of the device heterogeneity  analysis. $p_{loss}$ is set to \num{0.5} for both training and testing.} 
\begin{adjustbox}{width=0.7\columnwidth,center=\columnwidth}
\resizebox{\textwidth}{!}{
\centering
	\begin{tabular}{@{}l*{4}{S[table-format=-3.4]}@{}}
	\toprule
	& \multicolumn{2}{c}{\bfseries Training Data} & \multicolumn{2}{c}{\bfseries Testing Data (accuracy) } \\
	\cmidrule(lr){2-3} \cmidrule(lr){4-5}
	& {Phone 1} & {Phone 2} & {Phone 1} & {Phone 2} \\
	\cmidrule(lr){2-2} \cmidrule(lr){3-3} \cmidrule(lr){4-4} \cmidrule(lr){5-5}  
	\multirow{2}{*}{Proposed Method}  & {\checkmark} &   & 0.565 & 0.300 \\  
	&  &  {\checkmark} & 0.299 & 0.539  \\
	\midrule
	\multirow{2}{*}{$k$-NN}  & {\checkmark} &   & 0.395 & 0.235  \\  
	&  &  {\checkmark} & 0.230 & 0.391  \\ 
	\midrule
	\multirow{2}{*}{SVM}  & {\checkmark}  &   & 0.396 & 0.228  \\  
	&  &  {\checkmark} & 0.222 & 0.393 \\ 
	\bottomrule  
\end{tabular} 
} 
\end{adjustbox}
\label{tableDH}
\end{table}

\section{Conclusion}
\label{sec6}

This paper presented the design and evaluation of a novel cellular-based symbolic indoor positioning method. At its core, the proposed method utilizes DAEs to alleviate the effects randomized signal loss has on positioning. Experimental results demonstrated that the proposed method outperforms conventional methods with respect to several performance metrics. In its current form, the proposed method does not consider temporal information/patterns in location inference. Exploiting temporal dependencies among data samples helps maintain temporal coherence which could further enhance performance. In this respect, we intend to extend this work by integrating deep learning architectures suitable for processing time-series data (e.g., RNN Autoencoders \cite{10.5555/3045118.3045209}). 

\bibliographystyle{IEEEtran}
\bibliography{IEEEabrv,IEEEexample, Bibliography}

\end{document}